# RARD II: The 94 Million Related-Article Recommendation Dataset

Joeran Beel*[1, 2], Barry Smyth[3] and Andrew Collins*[1]

[1] Trinity College Dublin, School of Computer Science & Statistics, ADAPT Centre, Ireland
[2] National Institute of Informatics Tokyo, Digital Content and Media Sciences Division, Japan
[3] University College Dublin, Insight Centre for Data Analytics, Ireland

beelj@tcd.ie, barry.smyth@insight-centre.org, ancollin@tcd.ie

**Abstract.** The main contribution of this paper is to introduce and describe a new recommender-systems dataset (RARD II). It is based on data from Mr. DLib, a recommender-system as-a-service in the digital library and reference-management-software domain. As such, RARD II complements datasets from other domains such as books, movies, and music. The dataset encompasses 94m recommendations, delivered in the two years from September 2016 to September 2018. The dataset covers an item-space of 24m unique items. RARD II provides a range of rich recommendation data, beyond conventional ratings. For example, in addition to the usual (implicit) ratings matrices, RARD II includes the original recommendation logs, which provide a unique insight into many aspects of the algorithms that generated the recommendations. The logs enable researchers to conduct various analyses about a real-world recommender system. This includes the evaluation of meta-learning approaches for predicting algorithm performance. In this paper, we summarise the key features of this dataset release, describe how it was generated and discuss some of its unique features. Compared to its predecessor RARD, RARD II contains 64% more recommendations, 187% more features (algorithms, parameters, and statistics), 50% more clicks, 140% more documents, and one additional service partner (JabRef).

## 1 Introduction

The availability of large-scale, realistic, and detailed datasets is an essential element of many research communities, such as the information retrieval community (e.g. TREC [1,2], NTCIR [3,4], and CLEF [5,6]), the machine learning community (e.g. UCI [7], OpenML [8], KDD Cup [9]) and the recommender systems community. Particularly, the meta-learning and algorithm section community [10] as well as the automated machine-learning (AutoML) community [11] depend on large-scale datasets. Such datasets provide data for researchers to benchmark existing techniques, as well as to develop, train and evaluate new algorithms. They can also be essential when it comes to supporting the development of a research community. The recommender-systems

* This publication has emanated from research conducted with the financial support of Science Foundation Ireland (SFI) under Grant Number 13/RC/2106 and funding from the European Union and Enterprise Ireland under Grant Number CF 2017 0303-1.

community has been well-served by the availability of a number of datasets in popular domains including movies [12,13], books [14], music [15] and news [5,16]. The importance of these datasets is evidenced by their popularity among researchers and practitioners; for example, the MovieLens datasets have been downloaded 140,000 times in 2014 [12], and Google Scholar lists some 13,000 research articles and books that mention one or more of the MovieLens datasets [17].

The recommender-systems community has matured, and the importance of recommender systems has grown rapidly in the commercial arena. Researchers and practitioners alike have started to look beyond the traditional e-commerce / entertainment domains (e.g. books, music, and movies). However, there are few datasets that are suitable for meta-learning in the context of recommender systems, or suitable for research in the field of digital libraries. It is with this demand in mind that we introduce the RARD II dataset. RARD II is based on a digital-library *recommender-as-a-service* platform known as Mr. DLib [18–20]. Primarily, Mr. DLib provides related-article type recommendations based on a source / query article, to a number of *partner services* including the social-sciences digital library *Sowiport* [21–24] and the *JabRef* reference-management software [25].

The unique value of RARD II for recommender-systems and algorithm-selection research, stems from the scale and variety of data that it provides in the domain of digital libraries. RARD II includes data from 94m recommendations, originating from more than 13.5m queries. The dataset comprises two full years of recommendations, delivered between September 2016 and September 2018. Importantly, in addition to conventional ratings-type data, RARD II includes comprehensive recommendation logs. These provide a detailed account of the recommendations that were generated – not only the items that were recommended, but also the context of the recommendation (the source query and recommendation destination), and meta-data about the algorithms and parameters used to generate them. Compared to its predecessor RARD I [26] – which was downloaded 1,489 times between June 2017 and April 2019 – RARD II contains 64% more recommendations, 187% more features (algorithms, parameters, and statistics), 50% more clicks, 140% more documents, and JabRef as new partner.

In what follows, we describe this RARD II data release in more detail. We provide information about the process that generated the dataset and pay particular attention to a number of the unique features of this dataset.

## 2 Background / Mr. DLib

Mr. DLib is a recommendation-as-a-service (RaaS) provider [18]. It is designed to suggest related articles through partner services such as digital libraries or reference management applications. For example, Mr. DLib provides related-article recommendations to Sowiport to be presented on Sowiport's website alongside some source/target article (see **Fig. 1**). Sowiport was the largest social science digital library in Germany, with a corpus of 10m articles (the service was discontinued in December 2017).

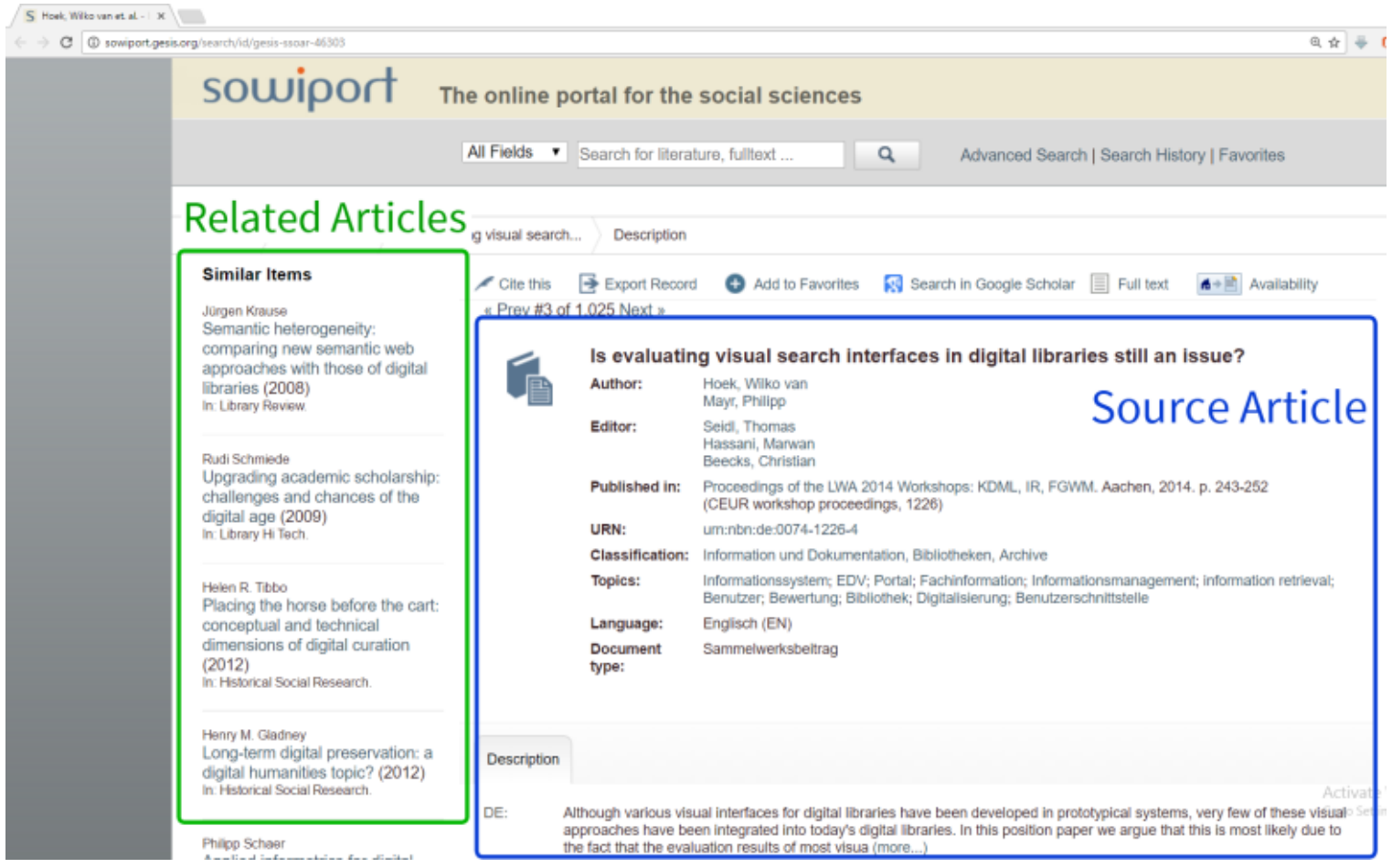

**Fig. 1.** Sowiport's website with a source-article (blue) and recommendations (green)

Fig. **2** summarises the recommendation process, implemented as a RESTful Web Service. The starting point for Mr. DLib to calculate recommendations is the ID or title of some source article. Importantly, Mr. DLib closes the recommendation loop because in addition to providing recommendations to a user, the response of a user (principally, article selections) is returned to Mr. DLib for logging.

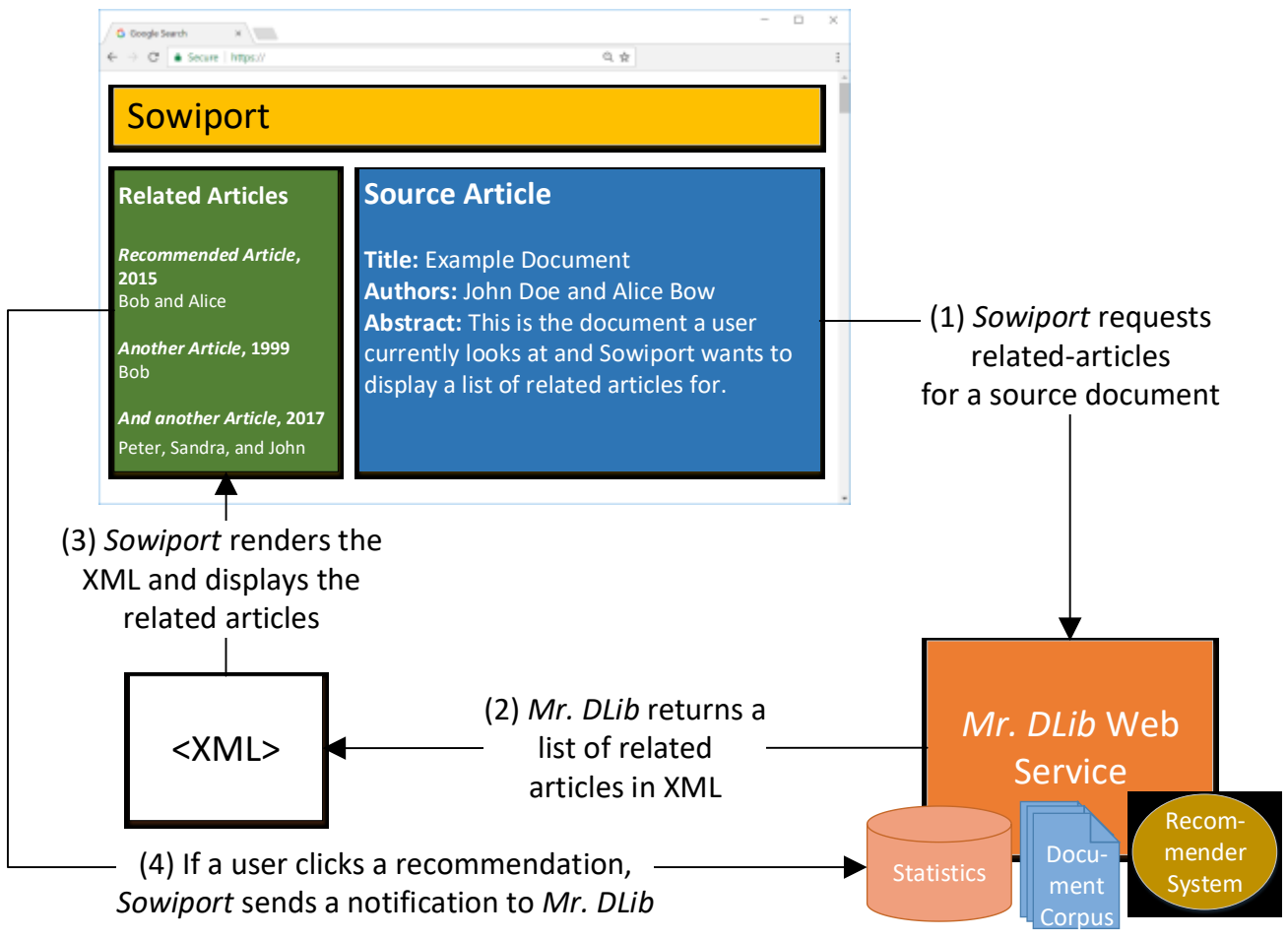

**Fig. 2.** Illustration of the recommendation process

In another use-case, Mr. DLib provides related-article recommendations to JabRef, one of the most popular open-source reference managers (**Fig. 3**) [27]. Briefly, when users select a reference/article in JabRef, the "related articles" tab presents a set of related-articles, retrieved from Mr. DLib. Related-article recommendations are made from the 10m Sowiport corpus, and 14m CORE documents [28–31].

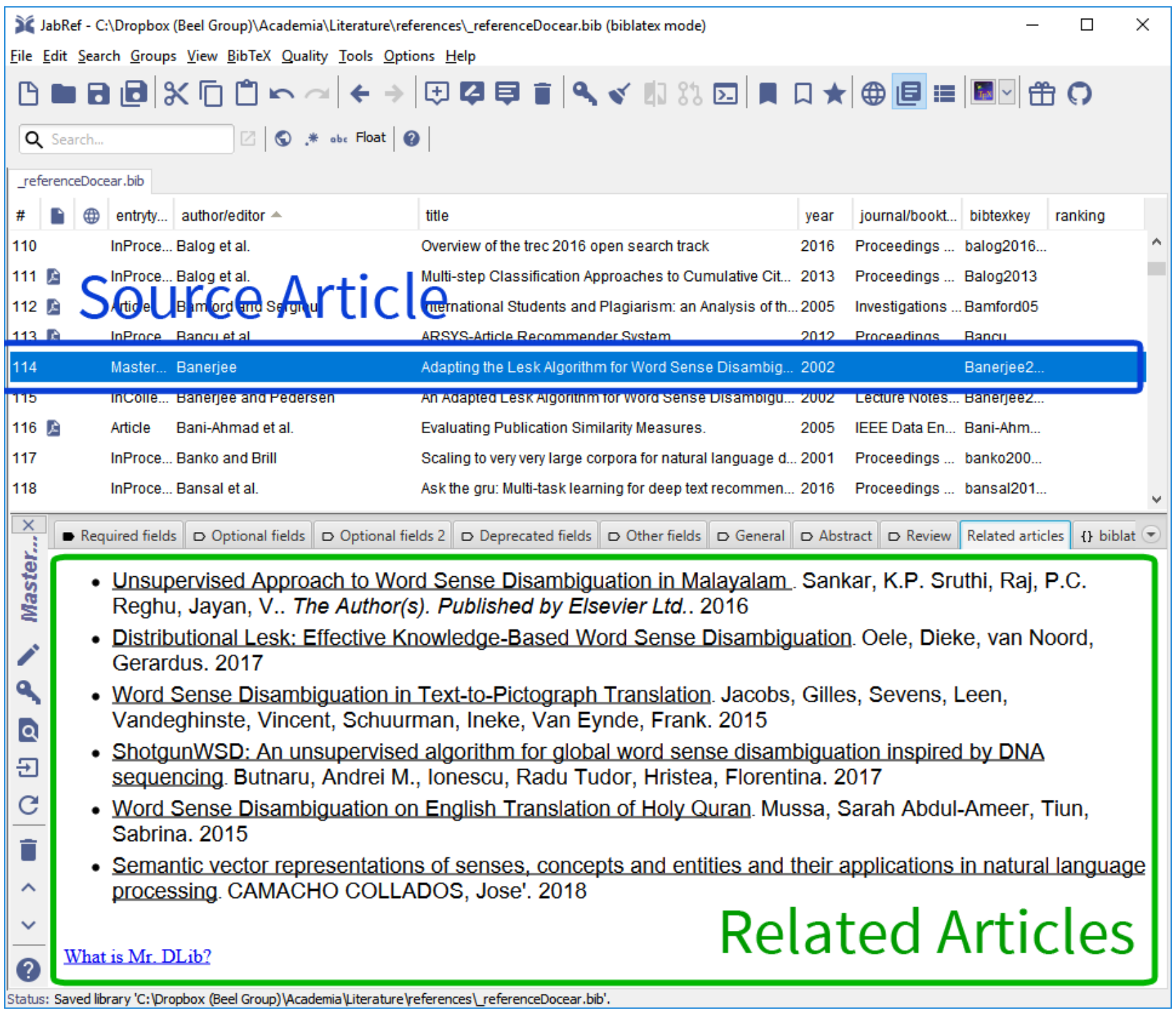

**Fig. 3.** Related-article recommendations in JabRef

To generate a set of recommendation, Mr. DLib harnesses different recommendation approaches including content-based filtering. Mr. DLib also uses external recommendation APIs such as the *CORE Recommendation API* [32,33] as part of a 'living lab' [34]. The algorithm selection and parametrization is managed by Mr. DLib's A/B testing engine. The details of the recommendation approach used, including any relevant parameters and configurations, are logged by Mr. DLib.

As an example of the data that Mr. DLib logs, when the A/B engine chooses content-based filtering, it randomly selects whether to use 'normal keywords', 'key-phrases' [35] or 'word embeddings'. For each option, additional parameters are randomly chosen; e.g., when key-phrases are chosen, the engine randomly selects whether key-phrases from the 'title' or 'abstract' are used. Subsequently, the system randomly selects whether unigrams, bigrams, or trigrams are used. Then, the system randomly selects how many key-phrases to use to calculate document relatedness, from one to twenty. The A/B engine also randomly chooses which matching mode to use when parsing queries [standardQP | edismaxQP]. Finally, the engine selects whether to re-rank recommendations with readership data from Mendeley, and how many recommendations to return.

All this information – the queries and responses, user actions, and the recommendation process meta data – is made available in the RARD II data release.

# 3 The RARD II Dataset

RARD II is available on http://data.mr-dlib.org and published under "*Creative Commons Attribution 3.0 Unported (CC-BY)*" license [36]. The dataset consists of three sub-datasets: (1) the r*ecommendation log*; (2) the *ratings matrices*; and (3) the *external document IDs*.

## 3.1 The Recommendation Log

The `recommendation_log.csv` file (20 GB) contains details on each related-article query from Sowiport and JabRef, and the individual article recommendations returned by Mr. DLib. A detailed description of every field in the recommendation log is beyond the scope of this paper (please refer to the dataset's documentation for full details). Briefly, the key statistics of the log are presented in **Table 1**.

**Table 1.** Key numbers of the recommendation log

| | Total | Sowiport | JabRef |
|---|---|---|---|
| **Requests** | | | |
| **Total** | 13,482,392 | 13,170,639 | 311,753 |
| **Unique (src_doc_id)** | 2,663,826[1] | 2,433,024 | 238,687 |
| **Responses** | | | |
| **Total** | 13,482,392 | 13,170,639 | 311,753 |
| **With 1+ Click(s)** | 103,856 | 100,578 | 3,278 |
| **Avg. #Recs per Response** | 6.97 | 6.99 | 5.92 |
| **Recommendations** | | | |
| **Total** | 93,955,966 | 92,110,708 | 1,845,258 |
| **Unique (rec_doc_id)** | 7,224,279[1] | 6,819,067 | 856,158 |
| **Clicks** | | | |
| **Total** | 113,954 | 110,003 | 3,951 |
| **Click-Through Rate** | 0.12% | 0.12% | 0.21% |

The recommendation log contains 93,955,966 rows (**Fig. 4**), and each row corresponds to a single recommended item, i.e. a related-article recommendation. All items were returned in one of the 13,482,392 responses to the 13,482,392 queries by Sowiport and JabRef. The 13.5m queries were made for 2,663,826 unique source documents (out of the 24m documents in the corpus). This means, for each of the 2.7m documents, recommendations were requested 5.2 times on average. For around 21.4m documents in the corpus, recommendations were never requested.

Each of the 13.5m responses contains between one and 15 related-article recommendations – 94m recommendations in total and 6.97 on average per response. The 94m recommendations were made for 7,224,279 unique documents out of the 24m documents in the corpus. This means, those documents that were recommended, were recommended 13 times on average. Around 17m documents in the corpus were never recommended.

[1] The sum of 'Sowiport' and 'JabRef' does not equal the 'Total' number because some documents were requested by / recommended to both Sowiport *and* JabRef. However, these documents are only counted once for the 'Total' numbers.

| row_id | query_id | query_received | partner | src_doc_id | rspns_id | rec_id | algo_id | text_field | … | rec_doc_id | re-ranking | responded | clicked |
|---|---|---|---|---|---|---|---|---|---|---|---|---|---|
| 1 | 1 | 18-Sep '16, 4:02 | sowiport | 5,265 | 1 | 1 | 239 | title | … | 95 | yes | 18-Sep '16, 4:03 | NULL |
| 2 | 2 | 18-Sep '16, 4:03 | sowiport | 854 | 2 | 2 | 21 | abstract | … | 4,588 | no | 18-Sep '16, 4:04 | NULL |
| 3 | 2 | 18-Sep '16, 4:03 | sowiport | 854 | 2 | 3 | 21 | abstract | … | 9,648 | no | 18-Sep '16, 4:04 | 18-Sep '16, 4:06 |
| 4 | 2 | 18-Sep '16, 4:03 | sowiport | 854 | 2 | 4 | 21 | abstract | … | 445 | no | 18-Sep '16, 4:04 | NULL |
| 5 | 3 | 18-Sep '16, 4:05 | sowiport | 917 | 3 | 5 | 3 | NULL | … | 776 | no | 18-Sep '16, 4:05 | 18-Sep '16, 4:08 |
| 6 | 3 | 18-Sep '16, 4:05 | sowiport | 917 | 3 | 6 | 3 | NULL | … | 95 | no | 18-Sep '16, 4:05 | NULL |
| … | … | … | … | … | … | … | … | … | … | … | … | … | … |
| 93,955,963 | 13,482,391 | 30-Sep '18, 23:48 | jabref | 5,265 | 13,482,391 | 93,955,963 | 21 | abstract | … | 95 | no | 30-Sep '18, 23:48 | 30-Sep '18, 23:48 |
| 93,955,964 | 13,482,391 | 30-Sep '18, 23:48 | jabref | 5,265 | 13,482,391 | 93,955,964 | 21 | abstract | … | 5,846 | no | 30-Sep '18, 23:48 | 30-Sep '18, 23:50 |
| 93,955,965 | 13,482,391 | 30-Sep '18, 23:48 | jabref | 5,265 | 13,482,391 | 93,955,965 | 21 | abstract | … | 778 | no | 30-Sep '18, 23:48 | NULL |
| 93,955,966 | 13,482,392 | 30-Sep '18, 23:50 | sowiport | 64 | 13,482,392 | 93,955,966 | 12 | title | … | 168 | yes | 30-Sep '18, 23:51 | NULL |

**Fig. 4.** Illustration of the recommendation log

For each recommended article, the recommendation log contains more than 40 features (columns) including:

- Information about the partner, query article, and time.
- The id of the recommendation response, recommended articles, and various ranking information before and after Mr. DLib's A/B selection.
- Information about the recommendation technique used, including algorithm features (relevancy and popularity metrics, content features where relevant) and the processing time needed to generate these recommendations.
- The user response (a click/selection, if any) and the time of this response.

The log includes 113,954 clicks received for the 94m recommendations, which equals a click-through rate (**CTR**) of 0.12%. Clicks were counted only once, i.e. if a user clicked the same recommendations multiple times, only the first click was logged. 103,856 of the 13.5m responses contained at least one clicked recommendation (0.77%). Based on feedback from colleagues and reviewers, we are aware that many believe a CTR of 0.12% would be very low as in some other recommender systems CTR is 5% and higher [37]. However, these other recommender systems provide personalized recommendations and display recommendations very prominently, for instance, in a pop-up dialog. Recommender systems that are like Mr. DLib – i.e. systems that provide non-personalized related-item recommendations in an unobtrusive way – achieve CTRs similar to Mr. DLib, or lower. The various user-interface factors that influence click-through rates are beyond the scope of this article. Suffice it to say that the Mr. DLib recommendations are typically presented in a manner that is designed not to distract the user, which no doubt tends to reduce the CTR.[2]

RARD II's recommendation log enables researchers to reconstruct the fine details of these historic recommendation sessions, and the responses of users, for the purpose of benchmarking and/or evaluating new article recommendation strategies, and training machine-learning models for meta-learning and algorithm selection.

---

[2] We recently re-implemented the system of Mr. DLib and observed higher click-through rates of around 0.7%. Presumably this may be caused by a much larger number of indexed items in the new Mr. DLib (120 million instead of 24 million). However, further analysis is necessary.

One example of an analysis would be the analysis how the effectiveness of the recommender system changes over time (**Fig. 5**). Between September 2016 and mid-February 2017, Mr. DLib delivered around 10 million recommendations per month to Sowiport. This number decreases to around 2 million recommendations per month from February 2017 onwards. This is due to a change in technology. In the beginning, Sowiport requested recommendations from their server also when web crawlers were crawling Sowiport's website. In February, Sowiport began to use a JavaScript client that was ignoring web crawlers. Click-through rate for Sowiport is slightly decreasing over time from around 0.17% in the beginning to around 0.14% in November 2017. Since December 2017, click-through rate on Sowiport decreased to near 0%. This decrease is due to Sowiport's termination of its main service, i.e. the search function, in December 2017. The search interface is deactivated, although the individual article's detail pages are still online. These pages are still indexed in Google and lead to some visitors on Sowiport. However, these visitors typically leave the website soon and click few or even no recommendations.

In April 2017, JabRef integrated Mr. DLib into its Version 4.0 beta (**Fig. 5**). The users of this beta version requested 1.7 thousand recommendations in April 2017. These numbers increased to 17 thousand recommendations in May and remained stable around 25 thousand recommendations in the following months until September 2017. Click-through rate during these months was comparatively high (up to 0.82%). Following the final release of JabRef Version 4.0 in October 2017, the volume of recommendations increased by factor 6, to an average of 150 thousand recommendations per month. The click-through rate stabilised around 0.2%.

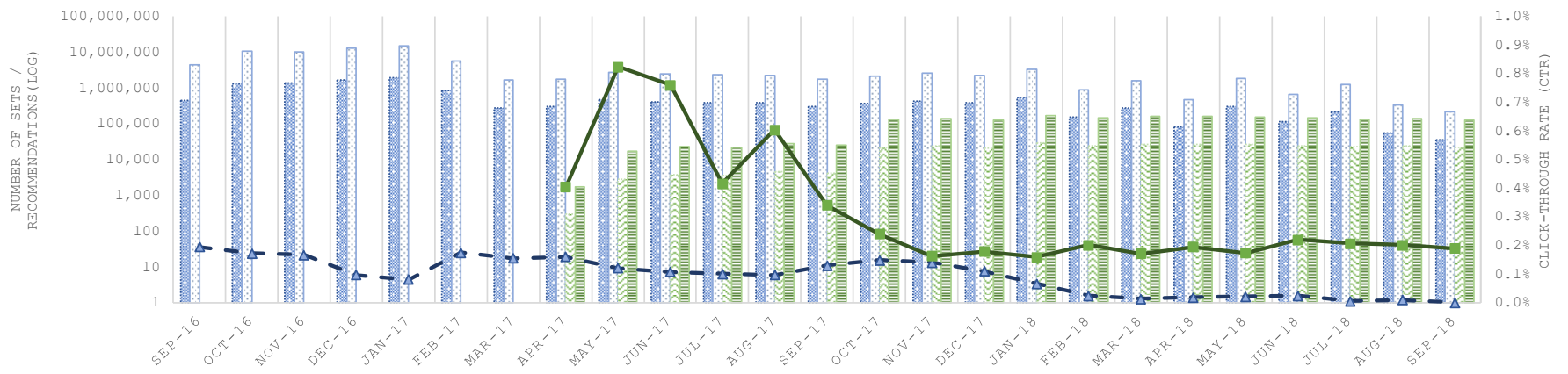

| | Sep-16 | Oct-16 | Nov-16 | Dec-16 | Jan-17 | Feb-17 | Mar-17 | Apr-17 | May-17 | Jun-17 | Jul-17 | Aug-17 |
|---|---|---|---|---|---|---|---|---|---|---|---|---|
| #Sets (Sowiport) | 440K | 1339K | 1392K | 1684K | 1986K | 834K | 280K | 304K | 462K | 416K | 392K | 386K |
| #Recs. (Sowiport) | 4.4M | 10.7M | 10.4M | 12.8M | 15.1M | 5.8M | 1.7M | 1.8M | 2.8M | 2.5M | 2.3M | 2.3M |
| #Sets (JabRef) | | | | | | | | 297 | 3K | 4K | 4K | 5K |
| #Recs. (JabRef) | | | | | | | | 1.7K | 16.9K | 22.4K | 21.7K | 27.5K |
| CTR (Sowiport) | 0.19% | 0.17% | 0.17% | 0.10% | 0.08% | 0.17% | 0.15% | 0.16% | 0.12% | 0.11% | 0.10% | 0.10% |
| CTR (JabRef) | | | | | | | | 0.40% | 0.82% | 0.76% | 0.42% | 0.60% |

| | Sep-17 | Oct-17 | Nov-17 | Dec-17 | Jan-18 | Feb-18 | Mar-18 | Apr-18 | May-18 | Jun-18 | Jul-18 | Aug-18 | Sep-18 |
|---|---|---|---|---|---|---|---|---|---|---|---|---|---|
| #Sets (Sowiport) | 301K | 361K | 432K | 383K | 551K | 151K | 271K | 80K | 308K | 112K | 213K | 56K | 36K |
| #Recs. (Sowiport) | 1.8M | 2.2M | 2.6M | 2.3M | 3.3M | .9M | 1.6M | .5M | 1.8M | .7M | 1.3M | .3M | .2M |
| #Sets (JabRef) | 4K | 22K | 24K | 21K | 29K | 24K | 27K | 27K | 26K | 24K | 23K | 23K | 21K |
| #Recs. (JabRef) | 24.7K | 131K | 142K | 125K | 173K | 144K | 160K | 160K | 153K | 144K | 135K | 139K | 126K |
| CTR (Sowiport) | 0.13% | 0.15% | 0.14% | 0.11% | 0.07% | 0.02% | 0.01% | 0.02% | 0.02% | 0.02% | 0.01% | 0.01% | 0.00% |
| CTR (JabRef) | 0.34% | 0.24% | 0.16% | 0.18% | 0.16% | 0.20% | 0.17% | 0.19% | 0.17% | 0.22% | 0.21% | 0.20% | 0.19% |

MONTH

**Fig. 5.** Number of Sets and Recommendations, and CTR by Month for Sowiport and JabRef

Further to content-based filtering recommendation approaches, Mr. DLib also recommends documents according to manually defined user-models with our stereotype algorithm [38]. We also recommend frequently viewed documents with our most-popular algorithm[38]. Content-based filtering algorithms are used most frequently, as our previous evaluations show that these are most effective for users of Sowiport and Jabref [19,38,39]. The distribution of algorithm usage within RARD II is shown in **Fig. 6**.

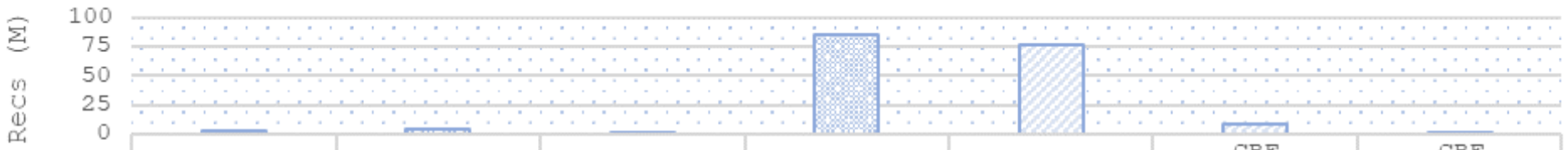

| | Random | Most Pop. | Stereotype | CBF (Total) | CBF (Terms) | CBF (Keyphrases) | CBF (Embeddings) |
|---|---|---|---|---|---|---|---|
| # Recs. | 3.1M | 4.5M | 1.6M | 84.8M | 76.0M | 8.5M | .3M |

**Fig. 6.** Total number of recommendations delivered with each algorithm

RARD II's detailed log entries make it uniquely suitable for analyses of algorithm selection approaches. Because all parameters for each algorithm are logged, as well as scenario attributes such as the time of day, the relationships between algorithm performance and this meta-data may be learned and evaluated. Meta-learning could be used to predict optimal algorithms per-request, using logged scenario attributes as meta-features. For example, a specific variant of content-based filtering might be most effective for users of Jabref at a certain time of day. Furthermore, as RARD II includes data from multiple recommendation scenarios, algorithm performance could be learned at a macro/global level, i.e., per platform [40].

Based on RARD II's recommendation logs, researchers could also analyse the performance of Mr. DLib's different recommendation algorithms and variations as shown in **Fig. 7**. The figure shows the CTR for the content-based filtering recommendations, stereotype recommendations, most-popular recommendations and a random recommendations baseline. In the first month, until February 2017, the CTR of all recommendation approaches is similar, i.e. between 0.1% and 0.2%. In March 2017, CTR for all approaches except content-based filtering decreases to near zero. We assume this to be due to the JavaScript that Sowiport used since March 2017, and which did not deliver recommendations to web spiders. Apparently, the comparatively high CTR of stereotype, most-popular and random recommendations was caused by web spiders following just all hyperlinks including the recommendations. As mentioned before, CTR for all recommendations, including content-based filtering, decreases again in December 2017 when Sowiport terminated its search feature.

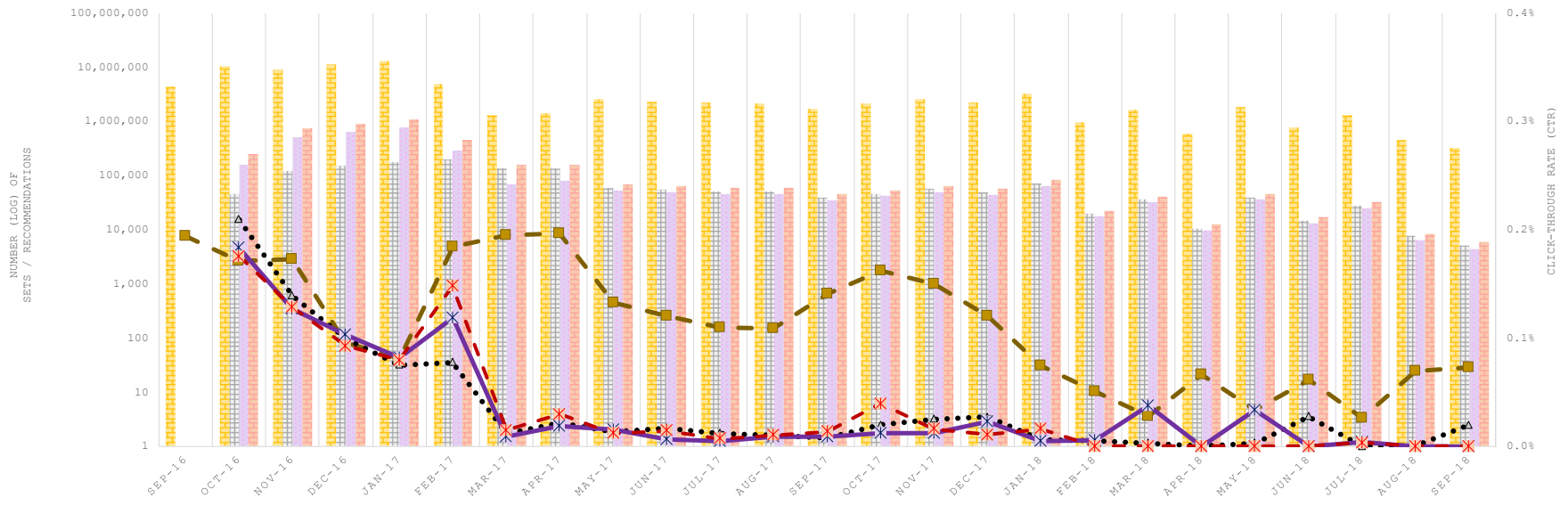

| | Sep-16 | Oct-16 | Nov-16 | Dec-16 | Jan-17 | Feb-17 | Mar-17 | Apr-17 | May-17 | Jun-17 | Jul-17 | Aug-17 |
|---|---|---|---|---|---|---|---|---|---|---|---|---|
| Cnt. Bsd. Flt., #Recs | 4M | 10M | 9M | 11M | 13M | 5M | 1M | 1M | 3M | 2M | 2M | 2M |
| Stereotype, #Recs | | 45K | 123K | 151K | 180K | 198K | 137K | 136K | 60K | 54K | 51K | 50K |
| Random, #Recs | | 157472 | 518140 | 641630 | 762873 | 288387 | 68052 | 81468 | 53934 | 48672 | 45672 | 45006 |
| Most Popular, #Recs | | 247K | 732K | 905K | 1073K | 452K | 156K | 157K | 68K | 63K | 58K | 59K |
| Cnt. Bsd. Flt., CTR | 0.19% | 0.17% | 0.17% | 0.10% | 0.08% | 0.18% | 0.19% | 0.20% | 0.13% | 0.12% | 0.11% | 0.11% |
| Stereotype, CTR | | 0.21% | 0.14% | 0.10% | 0.08% | 0.08% | 0.01% | 0.02% | 0.01% | 0.02% | 0.01% | 0.01% |
| Random, CTR | | 0.18% | 0.13% | 0.10% | 0.08% | 0.12% | 0.01% | 0.02% | 0.01% | 0.01% | 0.00% | 0.01% |
| Most Popular, CTR | | 0.18% | 0.13% | 0.09% | 0.08% | 0.15% | 0.01% | 0.03% | 0.01% | 0.01% | 0.01% | 0.01% |

| | Sep-17 | Oct-17 | Nov-17 | Dec-17 | Jan-18 | Feb-18 | Mar-18 | Apr-18 | May-18 | Jun-18 | Jul-18 | Aug-18 | Sep-18 |
|---|---|---|---|---|---|---|---|---|---|---|---|---|---|
| Cnt. Bsd. Flt., #Recs | 2M | 2M | 3M | 2M | 3M | 1M | 2M | .6M | 1.9M | .8M | 1.3M | .4M | .3M |
| Stereotype, #Recs | 39K | 46K | 56K | 49K | 72K | 20K | 36K | 10K | 40K | 15K | 28K | 8K | 5K |
| Random, #Recs | 34999 | 41622 | 49104 | 44310 | 64134 | 17538 | 31980 | 9516 | 36210 | 13302 | 24732 | 6444 | 4326 |
| Most Popular, #Recs | 46K | 54K | 64K | 57K | 83K | 23K | 41K | 12K | 46K | 17K | 33K | 8K | 6K |
| Cnt. Bsd. Flt., CTR | 0.14% | 0.16% | 0.15% | 0.12% | 0.07% | 0.05% | 0.03% | 0.07% | 0.03% | 0.06% | 0.03% | 0.07% | 0.07% |
| Stereotype, CTR | 0.01% | 0.02% | 0.02% | 0.03% | 0.01% | 0.01% | 0.00% | 0.00% | 0.00% | 0.03% | 0.00% | 0.00% | 0.02% |
| Random, CTR | 0.01% | 0.01% | 0.01% | 0.02% | 0.00% | 0.01% | 0.04% | 0.00% | 0.03% | 0.00% | 0.00% | 0.00% | 0.00% |
| Most Popular, CTR | 0.01% | 0.04% | 0.02% | 0.01% | 0.02% | 0.00% | 0.00% | 0.00% | 0.00% | 0.00% | 0.00% | 0.00% | 0.00% |

MONTH

**Fig. 7.** Number of Recommendations and CTR by Month for Content-Based Filtering, Stereotype Recommendations, Most-Popular Recommendations, and Random Recommendations.

The recommendation logs may also be used to research the effect that the feature type in content-based filtering has on performance (**Fig. 8**). Surprisingly, standard keywords perform best as shown in **Fig. 8**. For instance, CTR for key phrases and word embeddings is 0.04% and 0.05% respectively in October 2017. In contrast, CTR of standard terms is 0.16% in October.

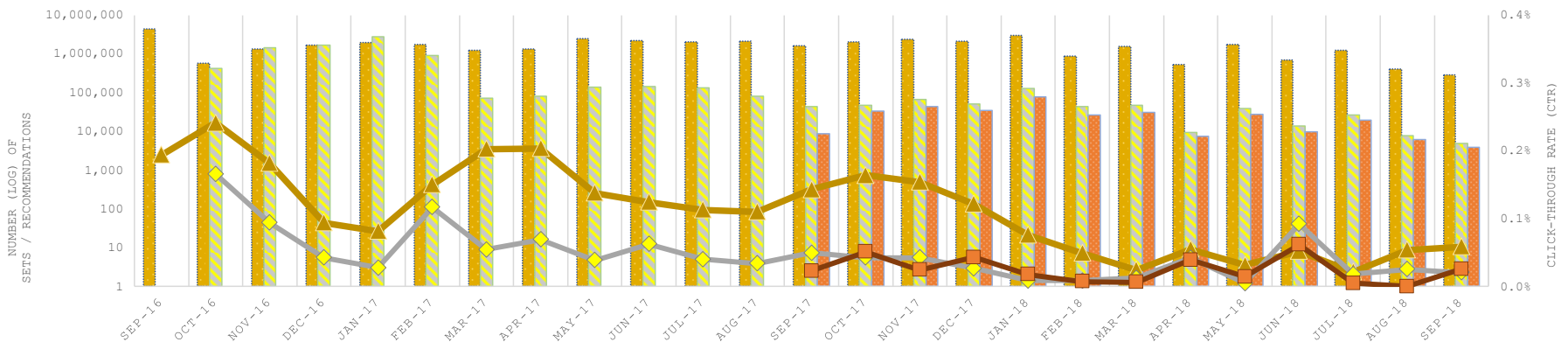

| | Sep-16 | Oct-16 | Nov-16 | Dec-16 | Jan-17 | Feb-17 | Mar-17 | Apr-17 | May-17 | Jun-17 | Jul-17 | Aug-17 |
|---|---|---|---|---|---|---|---|---|---|---|---|---|
| Keywords, #Recs | 4.40M | .58M | 1.36M | 1.68M | 2.00M | 1.72M | 1.22M | 1.35M | 2.46M | 2.20M | 2.07M | 2.09M |
| Keyphrases, #Recs | | 416K | 1M | 2M | 3M | 925K | 71K | 82K | 137K | 146K | 134K | 82K |
| Word Embd., #Recs | | | | | | | | | | | | |
| Keywords, CTR | 0.19% | 0.24% | 0.18% | 0.09% | 0.08% | 0.15% | 0.20% | 0.20% | 0.14% | 0.12% | 0.11% | 0.11% |
| Keyphrases, CTR | | 0.17% | 0.09% | 0.04% | 0.03% | 0.12% | 0.05% | 0.07% | 0.04% | 0.06% | 0.04% | 0.03% |
| Word Embd., CTR | | | | | | | | | | | | |

| | Sep-17 | Oct-17 | Nov-17 | Dec-17 | Jan-18 | Feb-18 | Mar-18 | Apr-18 | May-18 | Jun-18 | Jul-18 | Aug-18 | Sep-18 |
|---|---|---|---|---|---|---|---|---|---|---|---|---|---|
| Keywords, #Recs | 1.65M | 2.02M | 2.40M | 2.14M | 2.99M | .88M | 1.55M | .54M | 1.77M | .71M | 1.24M | .40M | .29M |
| Keyphrases, #Recs | 44K | 47K | 68K | 52K | 128K | 45K | 48K | 9K | 39K | 14K | 27K | 8K | 5K |
| Word Embd., #Recs | 9K | 33K | 45K | 35K | 79K | 27K | 31K | 8K | 28K | 10K | 19K | 6K | 4K |
| Keywords, CTR | 0.14% | 0.16% | 0.15% | 0.12% | 0.08% | 0.05% | 0.02% | 0.06% | 0.03% | 0.05% | 0.02% | 0.05% | 0.06% |
| Keyphrases, CTR | 0.05% | 0.04% | 0.04% | 0.03% | 0.01% | 0.01% | 0.01% | 0.04% | 0.01% | 0.09% | 0.02% | 0.03% | 0.02% |
| Word Embd., CTR | 0.02% | 0.05% | 0.02% | 0.04% | 0.02% | 0.01% | 0.01% | 0.04% | 0.01% | 0.06% | 0.01% | 0.00% | 0.03% |

MONTH

**Fig. 8.** Number of Recommendations and CTR by Month for Different Content-Based-Filtering Variations

RARD II also allows researchers to analyse the impact that processing time has on click-through rate. **Fig. 9** shows that the longer users need to wait for recommendations, the lower the click-through rate becomes. This holds true for all algorithms being used in Mr. DLib. For instance, when content-based filtering recommendations are returned within 2 second, the average CTR is 0.16%. If recommendations are returned after 5 seconds, CTR decreases to 0.11% and if recommendations are returned after 10 seconds, CTR decreases to 0.06%. This finding correlates with findings in other information retrieval applications such as search engines [41,42].

RARD II's recommendation log allows for many analyses more. Examples include or own analyses of the effect of bibliometric re-ranking [43], position bias (effect of a recommendation's rank, regardless of its actual relevance) [44], choice overload (effect of the number of displayed recommendations) [45], and the effect of the document field being used for content-based filtering [46].

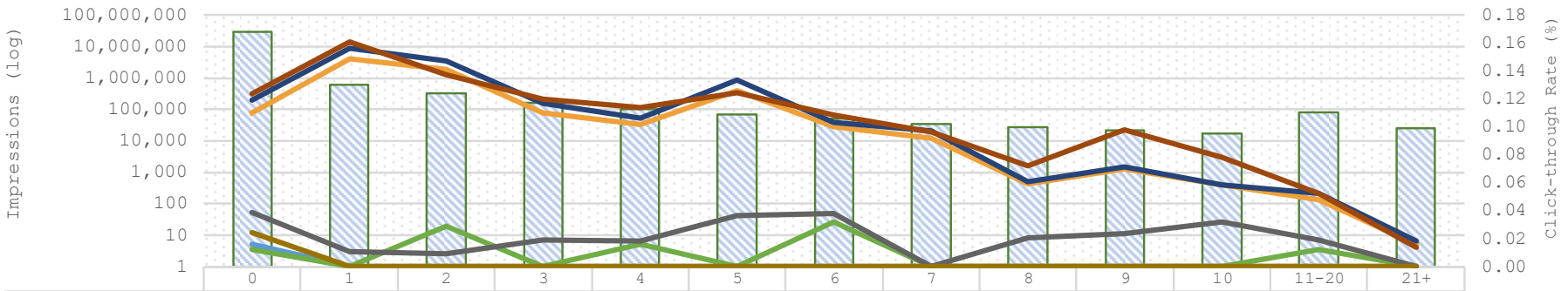

| | 0 | 1 | 2 | 3 | 4 | 5 | 6 | 7 | 8 | 9 | 10 | 11-20 | 21+ |
|---|---|---|---|---|---|---|---|---|---|---|---|---|---|
| Impressions | 29275k | 624k | 331k | 169k | 101k | 72k | 50k | 35k | 27k | 21k | 17k | 79k | 24k |
| All algorithms | 0.11 | 0.15 | 0.14 | 0.11 | 0.10 | 0.13 | 0.10 | 0.09 | 0.06 | 0.07 | 0.06 | 0.05 | 0.02 |
| Stereotype | 0.02 | 0.00 | 0.00 | 0.00 | 0.00 | 0.00 | 0.00 | 0.00 | 0.00 | 0.00 | 0.00 | 0.00 | 0.00 |
| Most Popular | 0.02 | 0.00 | 0.00 | 0.00 | 0.00 | 0.00 | 0.00 | 0.00 | 0.00 | 0.00 | 0.00 | 0.00 | 0.00 |
| Random | 0.01 | 0.00 | 0.03 | 0.00 | 0.02 | 0.00 | 0.03 | 0.00 | 0.00 | 0.00 | 0.00 | 0.01 | 0.00 |
| CBF | 0.12 | 0.16 | 0.15 | 0.12 | 0.11 | 0.13 | 0.10 | 0.10 | 0.06 | 0.07 | 0.06 | 0.05 | 0.02 |
| Terms | 0.12 | 0.16 | 0.14 | 0.12 | 0.11 | 0.12 | 0.11 | 0.10 | 0.07 | 0.10 | 0.08 | 0.05 | 0.01 |
| Keyphrases | 0.04 | 0.01 | 0.01 | 0.02 | 0.02 | 0.04 | 0.04 | 0.00 | 0.02 | 0.02 | 0.03 | 0.02 | 0.00 |
| Embeddings | 0.02 | 0.00 | 0.00 | 0.00 | 0.00 | 0.00 | 0.00 | 0.00 | 0.00 | 0.00 | 0.00 | 0.00 | 0.00 |

Processing Time (seconds)

Impressions · All algorithms · Stereotype · Most Popular · Random · CBF · Terms · Keyphrases · Embeddings

**Fig. 9.** CTR based on time to calculate recommendations

## 3.2 The Implicit Ratings Matrices

A ratings matrix is a mainstay of most recommendation datasets. RARD II contains two *implicit*, *item-item* rating matrices discussed below. Implicit ratings are based on the assumption that when users click a recommended article it is because they find the recommendation relevant (*a positive rating)*. And, conversely if they don't click a recommendation it is because the recommendation is not relevant (*a negative rating*). Of course, neither of these assumptions is particularly reliable. For example, just because a user selects a recommendation doesn't mean it will turn out to be relevant. Likewise, a user may choose not to respond if a recommendation is not relevant, or they may simply not notice the recommendation. However, click-related metrics such as click-through rate are a commonly used metric, particularly in industry, and are a good first indication of relevance.

The full ratings matrix `rating_matrix_full.csv` (1.8 GB) is illustrated in **Fig. 10**. It contains 48,879,170 rows, one for each <*source document; recommended document*> tuple. For each tuple, the following information is provided.

- The id of the source document (src_doc_id) for which Sowiport or JabRef queried recommendations. In a typical recommendation scenario, this entity may be interpreted as the user to whom recommendations were made.
- The id of a related-article (rec_doc_id) that was returned for the query. This entity may be interpreted as the item that was recommended to the 'user'.
- The number of times (#displayed) the tuple occurs in the recommendation log, i.e. how often the article was recommended as related to the given source document.
- The number of times the recommended article was clicked by a user (#clicks).
- The click-through rate (CTR), which represents an implicit rating of how relevant the recommended document was for the given source document.

| row_id | src_doc_id (user) | rec_doc_id (item) | # displayed | # clicks | ctr (rating) |
|---|---|---|---|---|---|
| 1 | 2 | 95 | 18 | - | 0% |
| 2 | 18 | 4,588 | 5 | 2 | 40% |
| 3 | 18 | 16,854,445 | 1 | - | 0% |
| 4 | 56 | 985 | 12 | 10 | 83% |
| ... | ... | ... | ... | ... | ... |
| 48,879,167 | 24,523,730 | 776 | 64 | 1 | 2% |
| 48,879,168 | 24,523,730 | 125,542 | 5 | - | 0% |
| 48,879,169 | 24,523,738 | 6,645 | 8 | - | 0% |
| 48,879,170 | 24,523,738 | 68,944 | 1 | 1 | 100% |

**Fig. 10.** Illustration of the full rating matrix

The full rating matrix was computed based on the full recommendation log. Hence, it also includes data from responses for which none of the recommendations were clicked. There are at least three reasons why users sometimes did not click on any recommendation: users may not have liked any of the recommendations; users may not have seen the recommendation; or recommendations may have been delivered to a web spider or bot that did not follow the hyperlinks. In the latter two cases, the non-clicks should not be interpreted as a negative vote. However, it is not possible to identify, which of the three scenarios applies for those sets in which no recommendation was clicked. Therefore, we created a filtered rating matrix.

The *filtered ratings* matrix **`rating_matrix_filtered.csv`** (26 MB) contains 745,167 rows and has the same structure (**Fig. 11**) as the full ratings matrix (**Fig. 10**). However, the matrix is based only on responses in which at least one recommendation from the response was clicked. The rationale is that when at least one recommendation was clicked, a real user must have looked at the recommendations who decided to click some recommendations and to not click others. Consequently, the non-clicked recommendations are more likely to correspond to an actual negative vote. Compared to the full matrix, the rows are missing that represent *<source document; recommended document>* tuples that were delivered in responses that did not receive any clicks. Also, the remaining rows tend to have lower *#displayed* counts than in the full-ratings matrix.

| id | src_doc_id (user) | rec_doc_id (item) | # displayed | # clicks | ctr (rating) |
|---|---|---|---|---|---|
| 1 | 2 | 95 | 13 | - | 0% |
| 2 | 18 | 4,588 | 2 | 2 | 100% |
| 3 | 56 | 985 | 11 | 10 | 91% |
| ... | ... | ... | ... | ... | ... |
| 745,165 | 24,523,730 | 776 | 32 | 1 | 3% |
| 745,166 | 24,523,730 | 125,542 | 2 | - | 0% |
| 745,167 | 24,523,738 | 68,944 | 1 | 1 | 100% |

**Fig. 11.** Illustration of the filtered rating matrix

There are certainly more alternatives to create the implicit ratings matrix. For instance, one may argue that recommendation sets in which all recommendations were clicked, might have been ‘clicked’ by a web spider, which simply followed all hyperlinks on a website. Based on the recommendation log, researchers can create their own implicit ratings matrix.

**Fig. 12** shows the distribution of views and clicks of the tuples (query document; recommended document). 93.9% of all tuples occur only once. This means, it rarely happened that a document pair (source document $x$ and recommended article $y$) was recommended twice or even more often. Actually, only 4.72% of the tuples were delivered twice, and 0.73% of the tuples were delivered three times. Most of these tuples did not receive any click (85.48%). 14% of the tuples received one click, and 0.4% received two clicks.

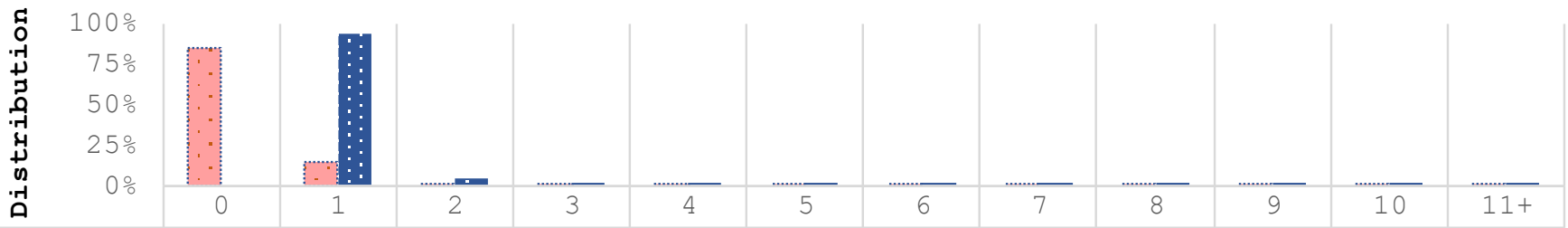

| | 0 | 1 | 2 | 3 | 4 | 5 | 6 | 7 | 8 | 9 | 10 | 11+ |
|---|---|---|---|---|---|---|---|---|---|---|---|---|
| Clicked | 85.440 | 14.187 | 0.279% | 0.051% | 0.019% | 0.008% | 0.005% | 0.002% | 0.002% | 0.001% | 0.001% | 0.007% |
| Displayed | | 94.64% | 4.12% | 0.70% | 0.23% | 0.11% | 0.06% | 0.04% | 0.02% | 0.02% | 0.01% | 0.05% |

Number of Views / Clicks

**Fig. 12.** Statistics of the filtered ratings matrix

### 3.3 External IDs (Sowiport, Mendeley, arXiv, …)

The third element of the data release is the list of external document ids (`external_IDs.csv`, 924 MB). There are 33m such ids for the Sowiport and CORE documents used. In addition, for a subset of the Sowiport ids there is associated identifiers for Mendeley (18%), ISSN (16%), DOI (14%), Scopus IDs (13%), ISBN (11%), PubMed IDs (7%) and arXiv IDs (0.4%). This third subset is provided to facilitate researchers in obtaining additional document data and meta-data from APIs provided by Sowiport [47], CORE [48], and Mendeley [49].

## 4 Discussion & Related Work

We have introduced RARD II, a large-scale, richly detailed dataset for recommender systems research based on more than 94m delivered scientific article recommendations. It contributes to a growing number of such datasets, most focusing on ratings data, ranging in scale from a few thousand datapoints to tens or even hundreds of millions of datapoints [50].

The domain of the RARD II data release (article recommendation) distinguishes it from more common e-commerce domains, but it is not unique in this regard [51–56]. Among others, CiteULike [57,58] and Bibsonomy [59,60] published datasets containing the social tags that their users added to research articles and, although not intended specifically for use in recommender systems research, these datasets have nonetheless been used to evaluate research-paper recommender systems [61–69]. Jack et al. compiled a Mendeley dataset [70] based on 50,000 randomly selected personal libraries from 1.5m users and with 3.6m unique articles. Similarly, Docear published a dataset based on its recommender system [71] based on the metadata of 9.4m articles, their citation network, related user information, and the details of 300,000 recommendations.

While RARD II shares some similarities with some of these datasets, particularly Docear and, of course, its predecessor RARD I [26], it is one of the only examples of a dataset from the digital library domain that has been specifically designed to support recommender systems research, and provides data at a scale that no other dataset in this domain provides. Because of this it includes information that is especially valuable to recommender systems researchers, not just ratings data but also the recommendation logs, which provide a unique insight into all aspects of the sessions that generated the recommendations and led to user clicks. Considering the scale and variety of data, RARD II is a unique and valuable addition to existing datasets.

Many datasets are pruned, i.e. data that is considered as not optimal is removed. For instance, the MovieLens datasets contain only data from users who rated at least 20 movies [12]. Such pruned datasets are nice for researchers because applying e.g. collaborative filtering works typically very well. However, such datasets are not realistic as dealing with noise is a crucial task in real-world recommender systems that are used in production. RARD II is not pruned, i.e. no data was removed. We believe that giving access to the full data including the many not-clicked recommendations is valuable for researchers who want to conduct research in a realistic scenario.

## 5 Limitations and Future Work

RARD II is a unique dataset with high value to recommender-systems researchers, particularly in the domain of digital libraries. However, we see areas for improvement, which we plan to address in the future.

Currently, RARD only contains the recommendation log and matrices from Mr. DLib, and Mr. DLib delivers recommendations only to two service partners. In the long run, we aim to make RARD a dataset that contains data from many RaaS operators. In addition to Mr. DLib, RaaS operators like Babel [72] and the CORE Recommender [33,48] could contribute their data. Additional service partners of Mr. DLib could also increase the value of the RARD releases. RARD would also benefit from having personalized recommendation algorithms included in addition to the non-personalized related-article recommendations. We are also aware of the limitations that clicks inherit as evaluation metrics. Future versions of RARD will include additional metrics such as real user ratings, and other implicit metrics. For instance, knowing whether a recommended document was eventually added to a JabRef user's library would provide valuable information. While RARD contains detailed information about the applied algorithms and parameters, information about the items is limited. We hope to be able to include more item-related data in future releases (e.g. metadata such as author names and document titles).

## References

[1] D. Harman, "Overview of the First Text REtrieval Conference (TREC-1)," *NIST Special Publication 500-207: The First Text REtrieval Conference (TREC-1)*, 1992.

[2] E.M. Voorhees and A. Ellis, eds., *The Twenty-Fifth Text REtrieval Conference (TREC 2016) Proceedings*, National Institute of Standards and Technology (NIST), 2016.

[3] A. Aizawa, M. Kohlhase, and I. Ounis, "NTCIR-10 Math Pilot Task Overview.," *NTCIR*, 2013.

[4] R. Zanibbi, A. Aizawa, M. Kohlhase, I. Ounis, G. Topi, and K. Davila, "NTCIR-12 MathIR task overview," *NTCIR, National Institute of Informatics (NII)*, 2016.

[5] F. Hopfgartner, T. Brodt, J. Seiler, B. Kille, A. Lommatzsch, M. Larson, R. Turrin, and A. Serény, "Benchmarking news recommendations: The clef newsreel use case," *ACM SIGIR Forum*, ACM, 2016, pp. 129–136.

[6] M. Koolen, T. Bogers, M. Gäde, M. Hall, I. Hendrickx, H. Huurdeman, J. Kamps, M. Skov, S. Verberne, and D. Walsh, "Overview of the CLEF 2016 Social Book Search Lab," *International Conference of the Cross-Language Evaluation Forum for European Languages*, Springer, 2016, pp. 351–370.

[7] D. Dheeru and E. Karra Taniskidou, "UCI Machine Learning Repository," *University of California, Irvine, School of Information and Computer Sciences. http://archive.ics.uci.edu/ml*, 2017.

[8] J. Vanschoren, J.N. van Rijn, B. Bischl, and L. Torgo, "OpenML: Networked Science in Machine Learning," *SIGKDD Explorations*, vol. 15, 2013, pp. 49–60.

[9] SIGKDD, "KDD Cup Archives," *ACM Special Interest Group on Knowledge Discovery and Data Mining. http://www.kdd.org/kdd-cup*, 2018.
[10] L. Kotthoff, "Algorithm selection for combinatorial search problems: A survey," *Data Mining and Constraint Programming*, Springer, 2016, pp. 149–190.
[11] F. Hutter, L. Kotthoff, and J. Vanschoren, "Automatic machine learning: methods, systems, challenges," *Challenges in Machine Learning*, 2019.
[12] F.M. Harper and J.A. Konstan, "The movielens datasets: History and context," *ACM Transactions on Interactive Intelligent Systems (TiiS)*, vol. 5, 2016, p. 19.
[13] S. Dooms, A. Bellogın, T.D. Pessemier, and L. Martens, "A Framework for Dataset Benchmarking and Its Application to a New Movie Rating Dataset," *ACM Trans. Intell. Syst. Technol.*, vol. 7, 2016, pp. 41:1–41:28.
[14] C.-N. Ziegler, S.M. McNee, J.A. Konstan, and G. Lausen, "Improving recommendation lists through topic diversification," *Proceedings of the 14th international conference on World Wide Web*, ACM, 2005, pp. 22–32.
[15] T. Bertin-Mahieux, D.P.W. Ellis, B. Whitman, and P. Lamere, "The Million Song Dataset," *Proceedings of the 12th International Conference on Music Information Retrieval (ISMIR 2011)*, 2011.
[16] B. Kille, F. Hopfgartner, T. Brodt, and T. Heintz, "The plista dataset," *Proceedings of the 2013 International News Recommender Systems Workshop and Challenge*, ACM, 2013, pp. 16–23.
[17] Google, "Search Results Mentioning MovieLens on Google Scholar," *https://scholar.google.de/scholar?q="movielens". Retrieved 14 April 2018*, 2018.
[18] J. Beel, A. Aizawa, C. Breitinger, and B. Gipp, "Mr. DLib: Recommendations-as-a-service (RaaS) for Academia," *Proceedings of the 17th ACM/IEEE Joint Conference on Digital Libraries*, Toronto, Ontario, Canada: IEEE Press, 2017, pp. 313–314.
[19] J. Beel and S. Dinesh, "Real-World Recommender Systems for Academia: The Gain and Pain in Developing, Operating, and Researching them," *Proceedings of the Fifth Workshop on Bibliometric-enhanced Information Retrieval (BIR) co-located with the 39th European Conference on Information Retrieval (ECIR 2017)*, P. Mayr, I. Frommholz, and G. Cabanac, eds., 2017, pp. 6–17.
[20] J. Beel, B. Gipp, S. Langer, M. Genzmehr, E. Wilde, A. Nuernberger, and J. Pitman, "Introducing Mr. DLib, a Machine-readable Digital Library," *Proceedings of the 11th ACM/IEEE Joint Conference on Digital Libraries (JCDL'11)*, ACM, 2011, pp. 463–464.
[21] D. Hienert, F. Sawitzki, and P. Mayr, "Digital library research in action - - supporting information retrieval in Sowiport," *D-Lib Magazine*, vol. 21, 2015.
[22] P. Mayr, "Sowiport User Search Sessions Data Set (SUSS)," *GESIS Datorium*, 2016.
[23] F. Sawitzki, M. Zens, and P. Mayr, "Referenzen und Zitationen zur Unterstützung der Suche in SOWIPORT," *Internationales Symposium der Informationswissenschaft (ISI 2013). Informationswissenschaft zwischen virtueller Infrastruktur und materiellen Lebenswelten*, DEU, 2013, p. 5.
[24] M. Stempfhuber, P. Schaer, and W. Shen, "Enhancing visibility: Integrating grey literature in the SOWIPORT Information Cycle," *International Conference on Grey Literature*, 2008, pp. 23–29.

[25] O. Kopp, U. Breitenbuecher, and T. Mueller, "CloudRef - Towards Collaborative Reference Management in the Cloud," *Proceedings of the 10th Central European Workshop on Services and their Composition*, 2018.

[26] J. Beel, Z. Carevic, J. Schaible, and G. Neusch, "RARD: The Related-Article Recommendation Dataset," *D-Lib Magazine*, vol. 23, Jul. 2017, pp. 1–14.

[27] J. Beel, "On the popularity of reference managers, and their rise and fall," *Docear Blog. https://www.docear.org/2013/11/11/on-the-popularity-of-reference-managers-and-their-rise-and-fall/*, 2013.

[28] P. Knoth and N. Pontika, "Aggregating Research Papers from Publishers' Systems to Support Text and Data Mining: Deliberate Lack of Interoperability or Not?," *Proceedings of INTEROP2016*, INTEROP2016, 2016.

[29] P. Knoth and Z. Zdrahal, "CORE: three access levels to underpin open access," *D-Lib Magazine*, vol. 18, 2012.

[30] N. Pontika, P. Knoth, M. Cancellieri, and S. Pearce, "Developing Infrastructure to Support Closer Collaboration of Aggregators with Open Repositories," *LIBER Quarterly*, vol. 25, Apr. 2016.

[31] L. Anastasiou and P. Knoth, "Building Scalable Digital Library Ingestion Pipelines Using Microservices," *Proceedings of the 11th International Conference on Metadata and Semantic Research (MTSR)*, Springer, 2018, p. 275.

[32] P. Knoth, L. Anastasiou, A. Charalampous, M. Cancellieri, S. Pearce, N. Pontika, and V. Bayer, "Towards effective research recommender systems for repositories," *Proceedings of the Open Repositories Conference*, 2017.

[33] N. Pontika, L. Anastasiou, A. Charalampous, M. Cancellieri, S. Pearce, and P. Knoth, "CORE Recommender: a plug in suggesting open access content," *http://hdl.handle.net/1842/23359*, 2017.

[34] J. Beel, A. Collins, O. Kopp, L. Dietz, and P. Knoth, "Mr. DLib's Living Lab for Scholarly Recommendations," *Proceedings of the 41st European Conference on Information Retrieval (ECIR)*, 2019.

[35] F. Ferrara, N. Pudota, and C. Tasso, "A Keyphrase-Based Paper Recommender System," *Proceedings of the IRCDL'11*, Springer, 2011, pp. 14–25.

[36] Creative Commons, "Creative Commons Attribution 3.0 Unported (CC BY 3.0)," *https://creativecommons.org/licenses/by/3.0/*, 2018.

[37] J. Beel, "Towards Effective Research-Paper Recommender Systems and User Modeling based on Mind Maps," *PhD Thesis. Otto-von-Guericke Universität Magdeburg*, 2015.

[38] J. Beel, S. Dinesh, P. Mayr, Z. Carevic, and J. Raghvendra, "Stereotype and Most-Popular Recommendations in the Digital Library Sowiport," *Proceedings of the 15th International Symposium of Information Science (ISI)*, 2017, pp. 96–108.

[39] S. Feyer, S. Siebert, B. Gipp, A. Aizawa, and J. Beel, "Integration of the Scientific Recommender System Mr. DLib into the Reference Manager JabRef," *Proceedings of the 39th European Conference on Information Retrieval (ECIR)*, 2017, pp. 770–774.

[40] A. Collins, D. Tkaczyk, and J. Beel, "A Novel Approach to Recommendation Algorithm Selection using Meta-Learning," *Proceedings of the 26th Irish Conference on Artificial Intelligence and Cognitive Science (AICS)*, CEUR-WS, 2018, pp. 210–219.

[41] R. Kohavi, A. Deng, R. Longbotham, and Y. Xu, "Seven rules of thumb for web site experimenters," *Proceedings of the 20th ACM SIGKDD international conference on Knowledge discovery and data mining*, ACM, 2014, pp. 1857–1866.
[42] E. Schurman and J. Brutlag, "The user and business impact of server delays, additional bytes, and HTTP chunking in web search," *Velocity Web Performance and Operations Conference*, 2009.
[43] S. Siebert, S. Dinesh, and S. Feyer, "Extending a Research Paper Recommendation System with Bibliometric Measures," *5th International Workshop on Bibliometric-enhanced Information Retrieval (BIR) at the 39th European Conference on Information Retrieval (ECIR)*, 2017.
[44] A. Collins, D. Tkaczyk, A. Aizawa, and J. Beel, "Position Bias in Recommender Systems for Digital Libraries," *Proceedings of the iConference*, Springer, 2018, pp. 335–344.
[45] F. Beierle, A. Aizawa, and J. Beel, "Choice Overload in Research-Paper Recommender Systems," *International Journal of Digital Libraries*, 2019.
[46] A. Collins and J. Beel, "Keyphrases vs. Document Embeddings vs. Terms for Recommender Systems: An Online Evaluation," *Proceedings of the ACM/IEEE-CS Joint Conference on Digital Libraries (JCDL)*, 2019.
[47] GESIS, "Sowiport OAI API," *http://sowiport.gesis.org/OAI/Home*, 2017.
[48] CORE, "CORE Open API and Datasets," *https://core.ac.uk*, 2018.
[49] Mendeley, "Mendeley Developer Portal (Website)," *http://dev.mendeley.com/*, 2016.
[50] A. Gude, "The Nine Must-Have Datasets for Investigating Recommender Systems," *Blog. https://gab41.lab41.org/the-nine-must-have-datasets-for-investigating-recommender-systems-ce9421bf981c*, 2016.
[51] M. Lykke, B. Larsen, H. Lund, and P. Ingwersen, "Developing a test collection for the evaluation of integrated search," *European Conference on Information Retrieval*, Springer, 2010, pp. 627–630.
[52] K. Sugiyama and M.-Y. Kan, "Scholarly paper recommendation via user's recent research interests," *Proceedings of the 10th ACM/IEEE Annual Joint Conference on Digital Libraries (JCDL)*, ACM, 2010, pp. 29–38.
[53] K. Sugiyama and M.-Y. Kan, "A comprehensive evaluation of scholarly paper recommendation using potential citation papers," vol. 16, 2015, pp. 91–109.
[54] K. Sugiyama and M.-Y. Kan, "Exploiting potential citation papers in scholarly paper recommendation," *Proceedings of the 13th ACM/IEEE-CS joint conference on Digital libraries*, ACM, 2013, pp. 153–162.
[55] D. Roy, K. Ray, and M. Mitra, "From a Scholarly Big Dataset to a Test Collection for Bibliographic Citation Recommendation," *AAAI Workshop: Scholarly Big Data*, 2016.
[56] D. Roy, "An improved test collection and baselines for bibliographic citation recommendation," *Proceedings of the 2017 ACM on Conference on Information and Knowledge Management*, ACM, 2017, pp. 2271–2274.
[57] CiteULike, "Data from CiteULike's new article recommender," *Blog, http://blog.citeulike.org/?p=136*, Nov. 2009.
[58] K. Emamy and R. Cameron, "Citeulike: a researcher's social bookmarking service," *Ariadne*, 2007.

[59] D. Benz, A. Hotho, R. Jäschke, B. Krause, F. Mitzlaff, C. Schmitz, and G. Stumme, "The Social Bookmark and Publication Management System BibSonomy," *The VLDB Journal*, vol. 19, 2010, pp. 849–875.

[60] Bibsonomy, "BibSonomy :: dumps for research purposes.," *https://www.kde.cs.uni-kassel.de/bibsonomy/dumps/*, 2018.

[61] C. Caragea, A. Silvescu, P. Mitra, and C.L. Giles, "Can't See the Forest for the Trees? A Citation Recommendation System," *iConference 2013 Proceedings*, 2013, pp. 849–851.

[62] R. Dong, L. Tokarchuk, and A. Ma, "Digging Friendship: Paper Recommendation in Social Network," *Proceedings of Networking & Electronic Commerce Research Conference (NAEC 2009)*, 2009, pp. 21–28.

[63] Q. He, J. Pei, D. Kifer, P. Mitra, and L. Giles, "Context-aware citation recommendation," *Proceedings of the 19th international conference on World wide web*, ACM, 2010, pp. 421–430.

[64] W. Huang, S. Kataria, C. Caragea, P. Mitra, C.L. Giles, and L. Rokach, "Recommending citations: translating papers into references," *Proceedings of the 21st ACM international conference on Information and knowledge management*, ACM, 2012, pp. 1910–1914.

[65] S. Kataria, P. Mitra, and S. Bhatia, "Utilizing context in generative bayesian models for linked corpus," *Proceedings of the 24th AAAI Conference on Artificial Intelligence*, 2010, pp. 1340–1345.

[66] D.M. Pennock, E. Horvitz, S. Lawrence, and C.L. Giles, "Collaborative filtering by personality diagnosis: A hybrid memory-and model-based approach," *Proceedings of the Sixteenth conference on Uncertainty in artificial intelligence*, Morgan Kaufmann Publishers Inc., 2000, pp. 473–480.

[67] L. Rokach, P. Mitra, S. Kataria, W. Huang, and L. Giles, "A Supervised Learning Method for Context-Aware Citation Recommendation in a Large Corpus," *Proceedings of the Large-Scale and Distributed Systems for Information Retrieval Workshop (LSDS-IR)*, 2013, pp. 17–22.

[68] R. Torres, S.M. McNee, M. Abel, J.A. Konstan, and J. Riedl, "Enhancing digital libraries with TechLens+," *Proceedings of the 4th ACM/IEEE-CS joint conference on Digital libraries*, ACM New York, NY, USA, 2004, pp. 228–236.

[69] F. Zarrinkalam and M. Kahani, "SemCiR - A citation recommendation system based on a novel semantic distance measure," *Program: electronic library and information systems*, vol. 47, 2013, pp. 92–112.

[70] K. Jack, M. Hristakeva, R.G. de Zuniga, and M. Granitzer, "Mendeley's Open Data for Science and Learning: A Reply to the DataTEL Challenge," *International Journal of Technology Enhanced Learning*, vol. 4, 2012, pp. 31–46.

[71] J. Beel, S. Langer, B. Gipp, and A. Nuernberger, "The Architecture and Datasets of Docear's Research Paper Recommender System," *D-Lib Magazine*, vol. 20, 2014.

[72] I. Wesley-Smith and J.D. West, "Babel: A Platform for Facilitating Research in Scholarly Article Discovery," *Proceedings of the 25th International Conference Companion on World Wide Web*, International World Wide Web Conferences Steering Committee, 2016, pp. 389–394.